\documentclass[fleqn,usenatbib]{mnras}
\usepackage[T1]{fontenc}

\usepackage{placeins}
\usepackage{graphicx}
\usepackage{dcolumn}
\usepackage{bm}

\usepackage{graphicx}
\usepackage{dcolumn}
\usepackage{bm}

\title[Mass Concentrations Search]{Searching for Mass Concentrations with Precision Pulsar Timing}
\author[John M.~LoSecco]{John M.~LoSecco$^{1}$\thanks{E-mail: losecco@nd.edu}
  \\
$^{1}$Physics and Astronomy Department, University of Notre Dame, Notre Dame, IN 46556-5670, USA}

\date{Accepted XXX. Received YYY; in original form ZZZ}

\pubyear{2025}

\begin{document}


\maketitle
\begin{abstract}
This papers searches for evidence of mass concentrations along the path of
radio pulses in the IPTA survey data release.
Radio pulse travel times are influenced via gravitational fields along the path
from the source to the observer.  Transient time delays in transit are a useful
measure of the matter distribution along the path.  Many pulsars have very
well understood timing solutions with predicable arrival times and can be used to sample
the mass variation.  Changes in the source, observer and mass concentration positions produce changes in arrival times which can be significant for precision pulsar times.
Nine candidates are reported from this search.  After red noise reduction the number of
candidates drops to two.  When the epoch at the encounter times are removed the candidates vanish. Delayed and early event candidates are found  at about the same rate.
There is no clear evidence for a signal of gravitational delays from this search.
\end{abstract}



\section{\label{sec:level1}Introduction}
Gravitational interactions have been suggested as a source of transient
distortions in pulsar arrival time. \cite{AA1995}.  Recently efforts have been
made to understand and correct for a broad class of timing distortions in
pulsar arrival times with results approaching microsecond resolution.
This suggests that searching these data may help strengthen models of galactic
mass distributions.

\begin{figure}
\includegraphics[width=\columnwidth]{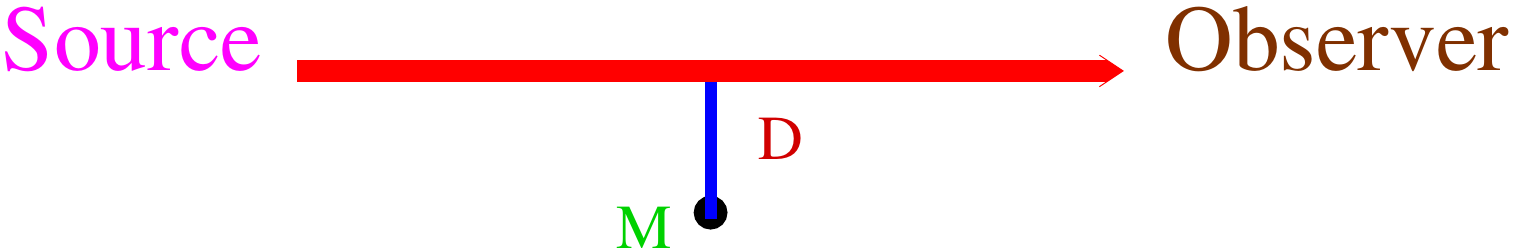}
  \caption{An illustration of the source to observer geometry.
  The impact parameter $D$ and the gravitating mass $\mathcal{M}$ are shown.}
  \label{fig:Samp_geom}
\end{figure}

In this note we search for the relativistic delay in pulsar
pulses due to encounters with the gravitational fields of masses distributed close the path of source to observer, as illustrated in figure \ref{fig:Samp_geom}.
A static mass distribution would add an unmeasurable constant time delay to the signal, so just as in other mass concentration searches \cite{Gaia} one looks for a time delay caused by a transient encounter.

The accumulated time delay is given by \cite{Wine,Larch}:
\begin{equation}
  \Delta T = -\frac{2G\mathcal{M}}{c^{3}} \ln( 1 - \hat{R} \cdot\hat{s})
  \label{eqn:ShapDel}
  \end{equation}
  where $\frac{2G\mathcal{M}}{c^{3}}$ is the Schwarzschild radius divided
by the speed of light and $\hat{R} \cdot\hat{s}$ is the cosine of the angle, as viewed from the observer, between the source of the pulse and the source of the gravitational field created by mass $\mathcal{M}$.

Pulsar timing arrays \cite{PPTA2} have improved arrival time fits for pulses and modeled the various known sources of time deviations in the transit time.  Orbital motion of the source and the observer can be fit and removed.  Observing at multiple frequencies permits removal of frequency dependent plasma delays.  Even the relativistic delay caused by a massive
companion of the pulsar can be fit and removed.

\section{\label{sec:Delay}Gravitational Time Delay}
\begin{figure}
\centering
\includegraphics[width=0.25\linewidth]{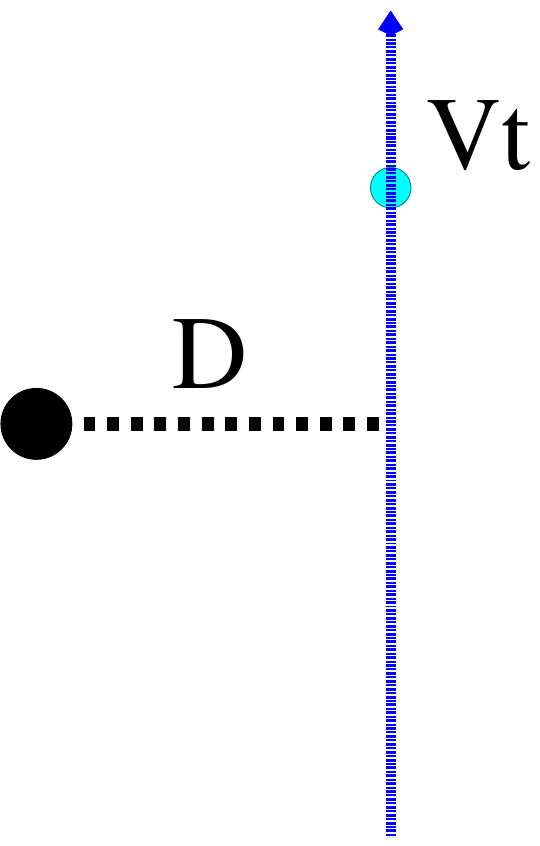}
	\caption{The time dependent geometry of the pulsar - mass concentration cosine as viewed by the observer. The Z axis goes from the observer to the pulsar (dot on the left). The point of closest approach is a distance $D$ along the X axis. The Y coordinate is upward along the projected speed of the mass concentration. The displacement along Y is taken as $Vt$.  The mass concentration is in the plane at (D,Vt).}
	\label{fig:ptdmgeom}
\end{figure}

In this paper we fit the pulsar timing residuals of the IPTA survey sources \cite{IPTA} to a 4 parameter model of an encounter of the beam with a mass source moving with constant speed in the plane of the observation. 

\cite{AA1995} suggest the 4 parameter fit of equation \ref{eqn:ptdmgeom}.
\begin{equation}
  \Delta T=-M~\ln{(1-\frac{1}{\sqrt{1+(v(t-t_{0}))^{2}+d^{2}}})}
  \label{eqn:ptdmgeom}
\end{equation}
with $M=\frac{2G\mathcal{M}}{c^{3}}$ as the mass of the source of the field
(in units of microseconds), $t_{0}$ is the time of the closest
encounter, $v$ is the (assumed) constant projected speed of the encounter
and $d$ is the distance of closest approach, at $t=t_{0}$.  For reference
$M_{sun}$ is 9.85 microseconds.  To be more precise
$d$ is the tangent of the angle of closest approach and $v(t-t_0)$ is the tangent
of the angle from the point of closest approach to the location at time $t$

Equation \ref{eqn:ptdmgeom} is found by explicitly calculating the value of $\hat{R} \cdot\hat{s}$ in the geometry shown in figure \ref{fig:ptdmgeom}.
Take the $\hat{z}$ axis along the direction from the observer to the pulsar.
The distance from the pulsar beam path to the mass concentration candidate point of closest approach is $D$.  The direction from
the pulsar beam path to the mass concentration candidate point of closest approach defines the $\hat{x}$ axis.  The vertical axis in the figure is the $\hat{y}$ axis and is the direction of motion of the mass concentration candidate.  The Z distance is arbitrary but we can take it as the distance along the $\hat{z}$ axis to the point of closest approach of the beam to the mass concentration candidate.
Figure \ref{fig:ptdmgeom} is drawn in the plane perpendicular to the $\hat{z}$ axis
at distance Z.  In this coordinate system $\vec{R}=(0,0,Z)$  and $\vec{s}=(D,Vt,Z)$.
\begin{equation}
\begin{array}{c}
\displaystyle
\hat{R} \cdot \hat{s}  = \frac{Z^2}{Z \times \sqrt{Z^2+(Vt)^2+D^2}}\\
\displaystyle
= \frac{Z^2}{Z^2 \sqrt{1+(vt)^2+d^2}}\\
\displaystyle
=\frac{1}{\sqrt{1+(vt)^2+d^2}}
\end{array}
\label{eqn:ptdmgeomder}
\end{equation}
which is equation \ref{eqn:ptdmgeom} with $d=D/Z$ and $v=V/Z$.

Equation \ref{eqn:ptdmgeomder} has some physical consequences.  $d$ is the tangent of an angle and $v$ is the time derivative of the tangent of an angle.  If the physical relative
velocity $V$ between the pulsar and the moving matter is roughly constant more distant encounters will have slower time delay encounters.  $d$ also decreases with greater distance which means that the maximum amplification
at closest approach will be larger for distant objects.  This is essentially parallax. Due to the smaller $v$ distant encounters may not return to baseline over the search window.  One may get overlapping encounters which will change the shape of the observed  delay.

\section{\label{sec:level2}The Data}
The IPTA release \cite{IPTA} includes 65 millisecond pulsars and their fits.  The Version B TCB data was processed with {\em Tempo2} \cite{Tempo2}, with and without maximum-likelihood red noise subtraction,
to provide residuals, the deviation of the pulse arrival time
from the time estimated by the model.  These residuals hold information about potential encounters with massive bodies
on route to the detector.

The IPTA data release 2 combines data from prior releases, EPTA 1,
NANOGrav 9-year, PPTA and their extensions.
The observations have been made by seven different radio telescopes (Eﬀelsberg,
Lovell, Nan\c{c}ay, Westerbork, Green Bank, Arecibo, Parkes)
The pulsar data samples span varying time intervals from 231 to 10753 days and
have from 116 to 17487 fitted pulses.  A total of 210148 pulse residuals over a
total time period of 255327 MJD are provided in the release.
\section{Method}
The method used has two parts.  The first part searches for signal delay
candidates over a limited portion of the recorded data.  Once found a local
fit is done on that candidate to recover the fit parameters.

Based on gravitational lensing results the encounters are expected to occur over a period
of weeks or months.  The data sample span periods from less than a year to almost 30 years.  Reliable fits also require many more data points to be fit compared to the four parameters to be measured.  The method adopted is to fit overlapping regions of at about 120
consecutive residuals, less if there is not enough data to provide 120 samples.
The initial fit is done with simulated annealing which searches the four dimensional parameter space.  This is followed by a Levenberg-Marquardt nonlinear fitter.  A modest event candidate is refit with the candidate time at the center of the interval.  When possible the time width of the fit is chosen to be the point where the peak has dropped to 10\% of its maximum height.  The peak position and width are based on the parameters extracted from the search step.  The fit is sensitive to the tails of the fit function so having two tails to fit helps reject noise.  If the candidate time is near either end of the data sample that tail is truncated.  Using a sample size large compared to the encounter time makes the fit more sensitive to fluctuations in the underlying tempo2 fit since the fit curve returns to
zero on both sides of an encounter.
\begin{table*}
\caption{\label{tab:results}There are 9 candidates after duplicates have been removed.
	Nine pulsars are represented.
	The first 4 columns give details of the fit sample including the time
	duration of the sample and the number of points fit.  The fifth column gives the RMS value of the residuals used in the fit as a measure of the noise level in the sample.  The last 4 columns give the result of the centered 10\% fit
	including the $\chi^2$ confidence of the fit, the mass fitted, the significance of the mass
	fit in $\sigma$'s and the maximum amplitude of the fit.  The RMS and the maximum
	amplitude are given in microseconds.  The masses are given in solar masses}
\begin{tabular}{|c|c|c|c|c|c|c|c|c|}
\hline
Pulsar & Sample & Duration & Count & RMS & Significance & $\mathcal{M}\pm\sigma_\mathcal{M}$ in $\mathcal{M}_{Sun}$ & $\mathcal{M}/\sigma_\mathcal{M}$ & MaxAmp \\
\hline
J1744-1134 & 94 & 57.8 & 441 & 2.481 & 0.9986 & 0.002497$\pm$0.0006352 & 3.93 & 1.03 \\
J0613-0200 & 23 & 239.4 & 421 & 4.752 & 0.9982 & 0.01244$\pm$0.003663 & 3.396 & 4.471 \\
J1600-3053 & 105 & 122.7 & 378 & 1.555 & 0.9153 & -0.005644$\pm$0.001714 & -3.292 & -1.645 \\
J1939+2134 & 226 & 70.8 & 347 & 45.71 & 0.8367 & -0.3258$\pm$0.03308 & -9.848 & -50.39 \\
J2145-0750 & 127 & 90.9 & 447 & 11.48 & 0.7004 & -0.02353$\pm$0.005579 & -4.218 & -14.49 \\
J1747-4036 & 25 & 80.8 & 492 & 6.215 & 0.2668 & 0.0335$\pm$0.005774 & 5.803 & 10.84 \\
J1955+2908 & 17 & 109.8 & 386 & 3.843 & 0.2096 & 0.01039$\pm$0.002965 & 3.503 & 4.319 \\
J1643-1224 & 3 & 226.7 & 107 & 6.98 & 0.2025 & 0.04918$\pm$0.00765 & 6.428 & 9.98 \\
J1824-2452 & 1 & 554.5 & 103 & 1.985 & 0.1607 & -0.05956$\pm$0.01748 & -3.406 & -3.237 \\
\hline
\end{tabular}
\end{table*}
To proceed from the search fit to the centered fit the search result must have
a log likelihood ratio test at the 95\% level.  The fit itself should have an
Anderson-Darling test for gaussian residuals at the 95\% level and the
absolute value
of the fitted mass, $|M|$ must be at least 2 time its $\sigma$.  Accepting
$M<-2\sigma$ is included to study potential background fluctuations
indicated by negative time delays with the same time structure expected from
a point mass.

The selection is tightened to $3 \sigma$ plus the requirement that $\chi^2/DOF$ probability is greater than 10\% is applied to the centered fit before it can be accepted as a Shapiro delay candidate.


\section{\label{sec:level2}Technical Details}
From equation \ref{eqn:ShapDel} it is clear that the fit can diverge as $\hat{R} \cdot \hat{s} \rightarrow 1$.  Since the
residual data is finite it is unlikely that the solution will fall on this point but the
fitter may still explore this region to approach the best fit.  Due to finite
precision of computer representations of numbers $1-\cos{\theta}$ may approach 0 more rapidly
due to loss of precision.  The fitter uses the ``log1p'' function of $-\cos{\theta}$ to delay
the round off error in the logarithm function evaluation.  At even smaller values of $\theta$
the approximation of $1-\cos{\theta} = \theta^{2}/2$ is used as the argument to the logarithm.
Eventually that is replaced with $\ln(\theta^{2}/2 )= 2 \ln(\theta)-\ln(2)$.  This
provides access to a maximum amplification of about 1490 with standard double
precision calculations.

The transition from search to fit uses the preliminary fitted velocity from
the search to determine the temporal range of the fit.  Ideally one wants a
time range in which the fit is dominated by the signal and not the asymptotic
tails which should go to zero.  The first attempt is to construct a time
symmetric region about the candidate time with the amplitude dropping to 10\%
of the peak height.
In the case of a very local fluctuation the
search fitted velocity is very large and one ends up with very few timing
residuals in the narrow time range.  To get more statistics the refit region
is expanded to get at least 100 points.

Alternatively a systematic shift in the
residuals raises the baseline.  There is no baseline parameter in the fit
so this is turned into a very small velocity.  The small velocity results in
a very large time range for the refit, sometimes including the whole data sample
for the pulsar.  The refitting for nearby search regions then have a large
overlap, reducing their independence.  In this case the refit is shortened to
under 500 residuals.

In both cases the refit sample is centered on the candidate
to the extent possible and the initial values for the refit parameters are those
taken from the search.  ``to the extent possible'' is needed since the data
itself is not uniformly sampled in time
and is a merger of data from multiple instruments.  The search time is
put in the middle of the refit sample.  Gaps in the data sample and different
sampling rates may cause the centered fit to not have the same number of residuals before and after the event time.
Lower or upper time limits of the whole data sample also restrict the range
for the centered fit.

In the case of low velocities this adjustment procedure may
produce a small bias against a true
signal with the direction of proper motion close to the line of sight.
On the other hand the search samples may have a non-zero offset, perhaps coming from very long period signals.  A symptom of such an offset is a very small value
for $v$.  If $v \Delta t<10$ mrad the offset is removed by subtracting the sample mean from the sample.  $\Delta t$ is the actual time length of the fit sample.  This baseline adjustment was not need for any of the candidate
encounters in this paper.

An MCMC based search was used to explore the parameter space near the solution to improve the error estimates.  A fit was accepted as a candidate
if $\mathcal{M}/\sigma_{\mathcal{M}}>3$ and the $\chi^2/DOF$ probability exceed 0.1.

\section{\label{sec:level2}Results}
\begin{table*}
\caption{\label{tab:summary}Summary of the fit parameters for the 9 events}
\begin{tabular}{|c|c|c|c|c|c|}
\hline
Pulsar & Sample & $\mathcal{M}\pm\sigma_{\mathcal{M}}$ & $d\pm\sigma_{d}$ & $v\pm\sigma_{v}$ & $T_{0}\pm\sigma_{T_{0}}$ \\
\hline
J1744-1134 & 94 & 0.002497$\pm$0.0006352 & -0.00000006618$\pm$0.0000006069 & 7.024$\pm$10.37 & 55731$\pm$0.0000025036 \\
J0613-0200 & 23 & 0.01244$\pm$0.003663 & 0.0000009747$\pm$0.03935 & 1.071$\pm$1.031 & 53974$\pm$0.6218 \\
J1600-3053 & 105 & -0.005644$\pm$0.001714 & -0.00003053$\pm$213.8 & 5.283$\pm$5.063 & 56068$\pm$1.3192 \\
J1939+2134 & 226 & -0.3258$\pm$0.03308 & 0.0316$\pm$0.024 & 0.02235$\pm$0.01407 & 56630$\pm$3.7714 \\
J2145-0750 & 127 & -0.02353$\pm$0.005579 & 2.171E-12$\pm$0.0000001042 & -0.000006982$\pm$0.00002607 & 56487$\pm$2.0464 \\
J1747-4036 & 25 & 0.0335$\pm$0.005774 & 0.000006053$\pm$0.0000518 & 833$\pm$1545 & 56354$\pm$0.0000010407 \\
J1955+2908 & 17 & 0.01039$\pm$0.002965 & -0.00000005551$\pm$0.0000005509 & 0.4187$\pm$0.7247 & 56426$\pm$0.000045784 \\
J1643-1224 & 3 & 0.04918$\pm$0.00765 & 0.002731$\pm$0.005094 & 1.669$\pm$1.257 & 53292$\pm$0.082416 \\
J1824-2452 & 1 & -0.05956$\pm$0.01748 & 5.143$\pm$0.1742 & 4.426$\pm$1.654 & 53841$\pm$5.7474 \\
\hline
\end{tabular}
\end{table*}
The initial search was done on data with dispersion measure (DM) corrections
but not red noise subtraction.
 1833 candidates pass the search criteria to be recentered and refit.
 After the refit 10 of these have $|\mathcal{M}/\sigma_{\mathcal{M}}|>3$.
 (37 of these have $|\mathcal{M}/\sigma_{\mathcal{M}}|>2$).
All 10 of these have a fit with a $\chi^2/DOF$ probability $>0.1$,
the criteria in the centered fit to become a final event candidate.
The results of this search are shown in table \ref{tab:summary}.
Since the search uses overlapping regions and the centered fit can extend the
fit into regions already fit there are duplicate candidates in some cases.
Since the data samples do not overlap completely the duplicates are not
identical but the date is very close and the other parameters agree
to fractions of a $\sigma$.  One duplicate (J1747-4036) has been removed from the table.
\begin{table*}
\caption{\label{tab:sample}Summary of the data sample of the 9 pulsars}
\begin{tabular}{|c|c|c|c|}
\hline
Pulsar & Sample Length (MJD) & Sample Size & Mean Residual \\
\hline
J1744-1134 & 7262.00 & 9834 & 9.39E-18 \\
J0613-0200 & 5863.88 & 9322 & -1.22E-17 \\
J1600-3053 & 4493.04 & 9006 & 6.31E-18 \\
J1939+2134 & 10753.43 & 13659 & 3.20E-15 \\
J2145-0750 & 7243.47 & 8456 & 1.08E-16 \\
J1747-4036 & 609.33 & 2771 & -8.21E-17 \\
J1955+2908 & 2967.89 & 1459 & 0.00380 \\
J1643-1224 & 7356.24 & 8136 & -5.59E-17 \\
J1824-2452 & 2063.30 & 276 & 0 \\
\hline
\end{tabular}
\end{table*}
\begin{figure}
\includegraphics[width=\columnwidth,height=0.27\paperheight]{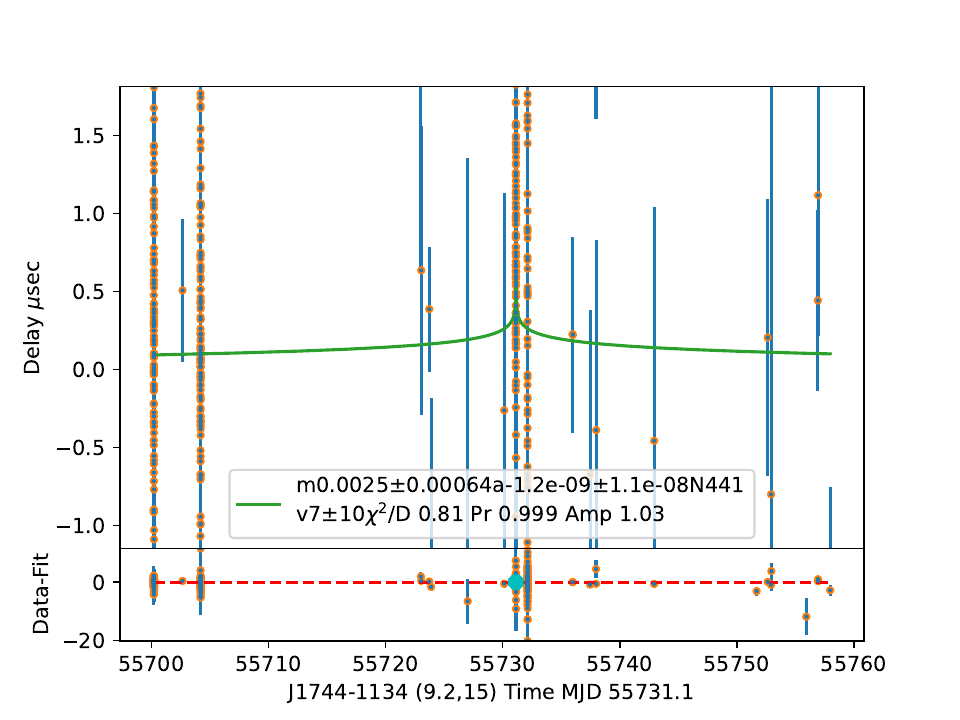}
\caption{\label{fig:DMbest1}The fit for J1744-1134 in the DM subtracted sample}
\end{figure}
\begin{figure}
\includegraphics[width=\columnwidth,height=0.27\paperheight]{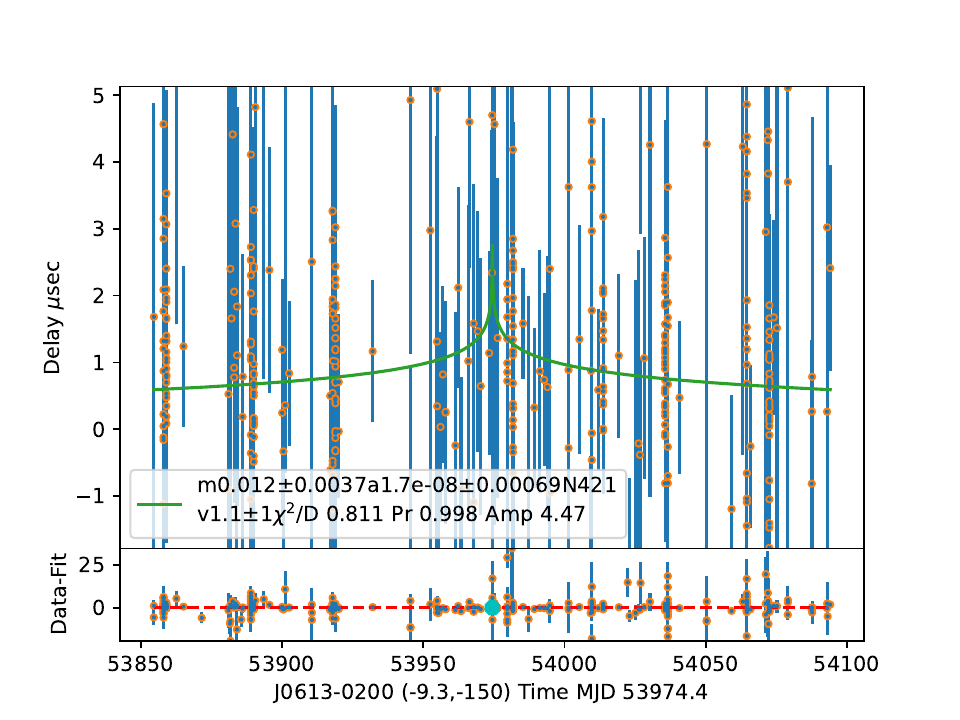}
\caption{\label{fig:DMbest2}The fit for J0613-0200 in the DM subtracted sample}
\end{figure}
\begin{figure}
\includegraphics[width=\columnwidth,height=0.27\paperheight]{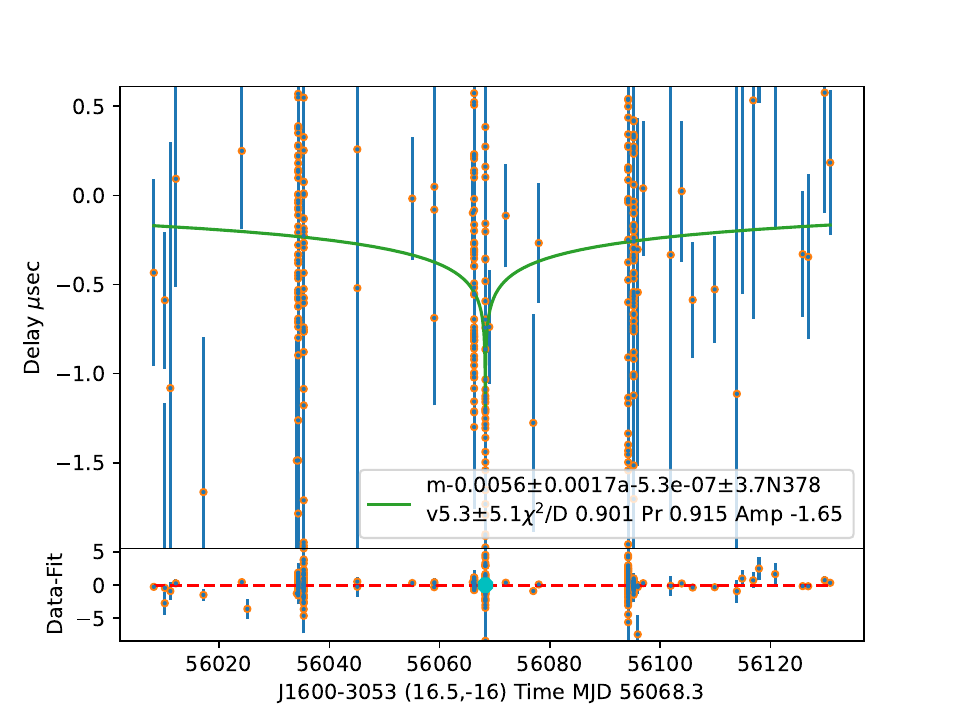}
\caption{\label{fig:DMbest3}The fit for J1600-3053 in the DM subtracted sample}
\end{figure}
Table \ref{tab:results} lists the results of the search.
The table includes details about the sample and results of the fit including
the fitted mass and the $\chi^2/DoF$ probability.  The number of residuals fit,
the duration of the sample (in MJD) and the RMS of the sample and the maximum amplitude of the fit, in microseconds, are also listed 

Table \ref{tab:summary} includes the full list of the fitted parameters
including $d$, $v$ and the time of closest approach $t_0$.   Note that there is not necessarily a data point taken at closest approach.  There are no
selection constraints on the values or significance of these other parameters.
These data are also printed on the plots.

\section{\label{sec:level2}Figures}
\begin{figure}
\includegraphics[width=\columnwidth,height=0.27\paperheight]{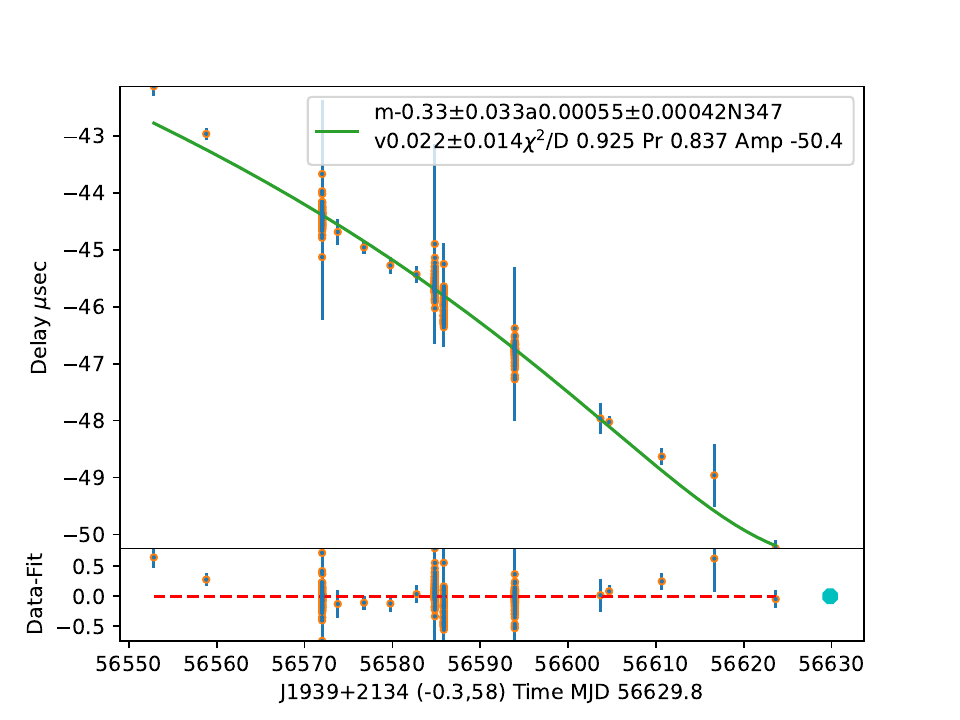}
\caption{\label{fig:DMbest4}The fit for J1939+2134 in the DM subtracted sample}
\end{figure}

\begin{figure}
\includegraphics[width=\columnwidth,height=0.27\paperheight]{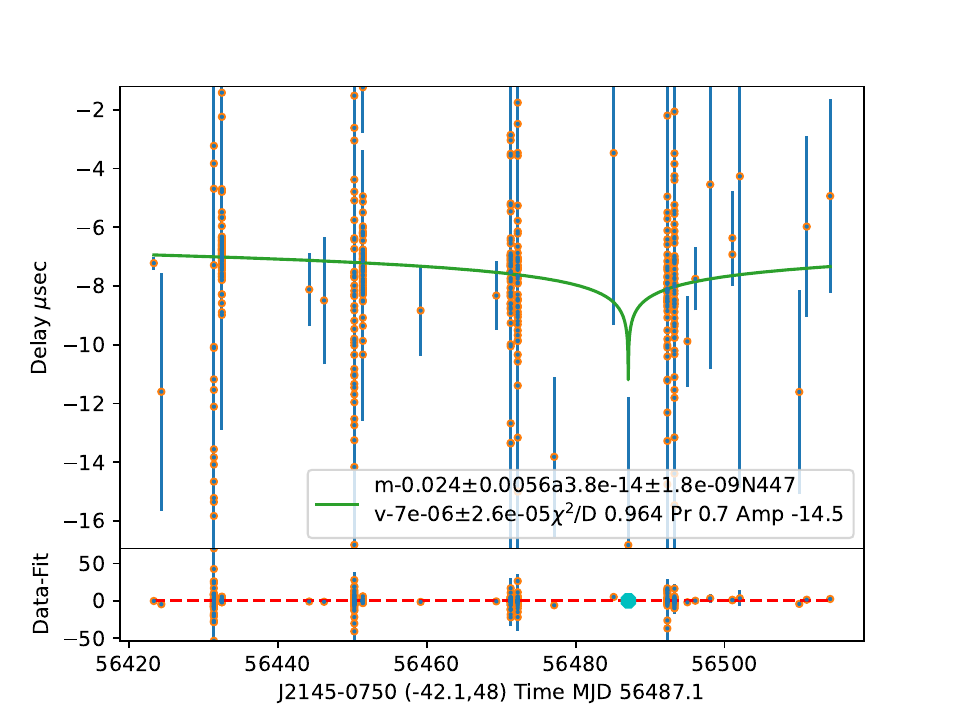}
\caption{\label{fig:DMbest5}The fit for J2145-0750 in the DM subtracted sample}
\end{figure}

\begin{figure}
\includegraphics[width=\columnwidth,height=0.27\paperheight]{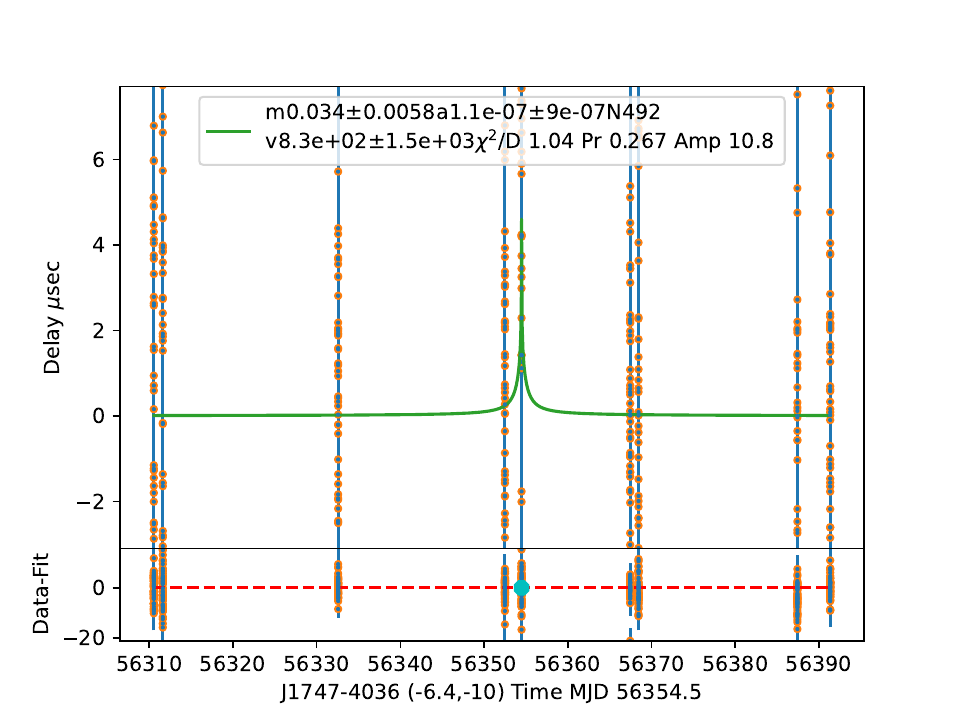}
\caption{\label{fig:DMbest6}The fit for J1747-4036 in the DM subtracted sample}
\end{figure}

\begin{figure}
\includegraphics[width=\columnwidth,height=0.27\paperheight]{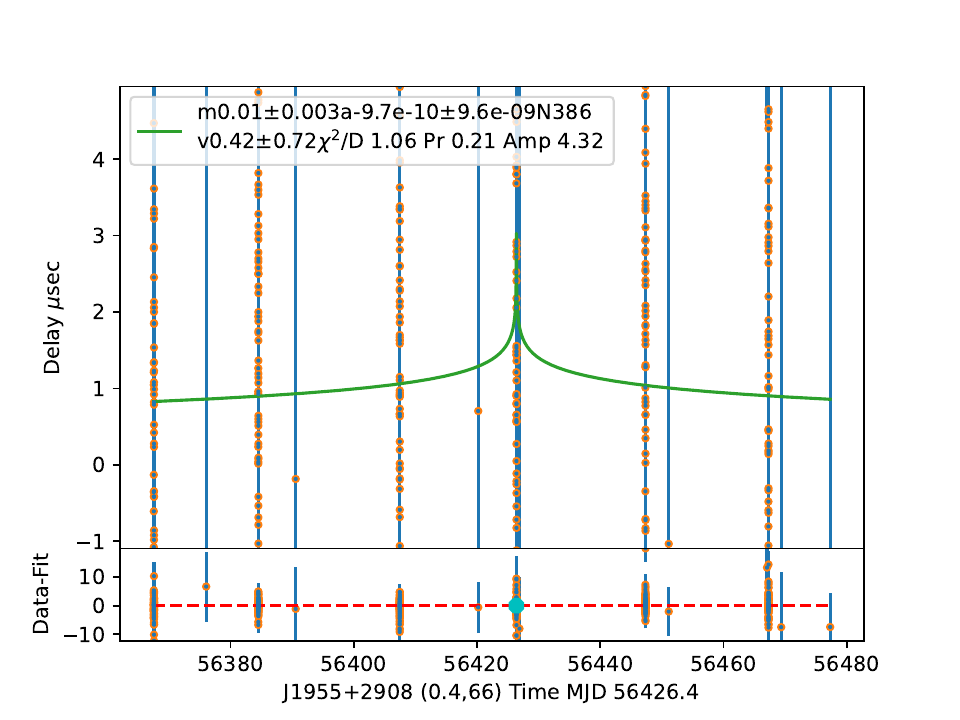}
\caption{\label{fig:DMbest7}The fit for J1955+2908 in the DM subtracted sample}
\end{figure}

\begin{figure}
\includegraphics[width=\columnwidth,height=0.27\paperheight]{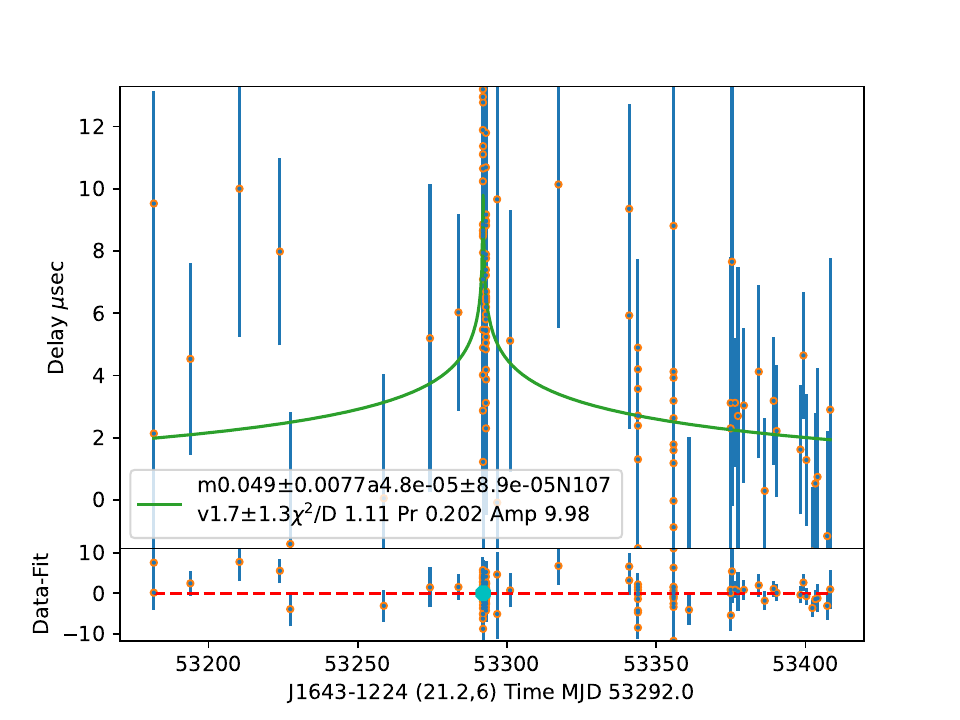}
\caption{\label{fig:DMbest8}The fit for J1643-1224 in the DM subtracted sample}
\end{figure}

\begin{figure}
\includegraphics[width=\columnwidth,height=0.27\paperheight]{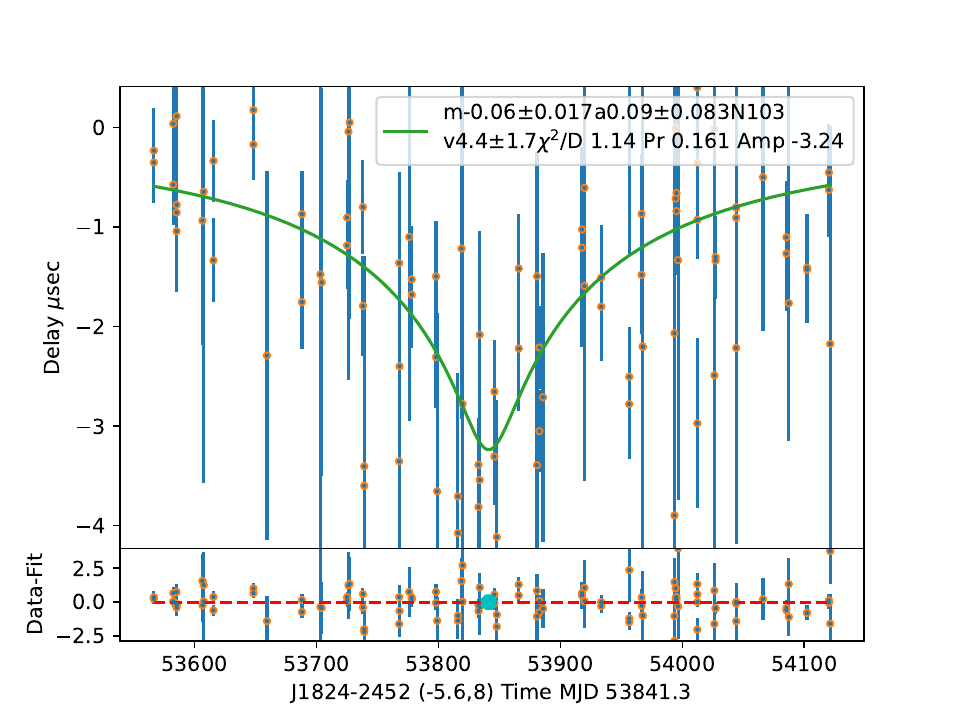}
\caption{\label{fig:DMbest9}The fit for J1824-2452 in the DM subtracted sample}
\end{figure}

Figures showing all the fits are displayed in Figures \ref{fig:DMbest1} to \ref{fig:DMbest9}.  The figures are ordered from high to low by $\chi^2/DOF$ probability.
Each figure has an upper panel showing the data and the fitted curve.  The
lower panel
shows the data minus the fitted value of the delay curve, the new residuals.
The dashed line on the lower panel is zero and the large point marks the
position of the peak
in the upper plot.  The vertical axes on each plot are individually scaled
and range from microseconds to tens of microseconds.
Values for ``vel''=$v$ and ``ang''=$a$ printed on the plots are scaled by
10,000 to reduce the number of zeros to the right of the decimal.  The plot legend also
includes the number of points fit (N), the $\chi^2/DOF$, the $\chi^2$ probability and
the maximum amplitude of the fitted curve (Amp).
\begin{figure}
\includegraphics[width=\columnwidth,height=0.28\paperheight]{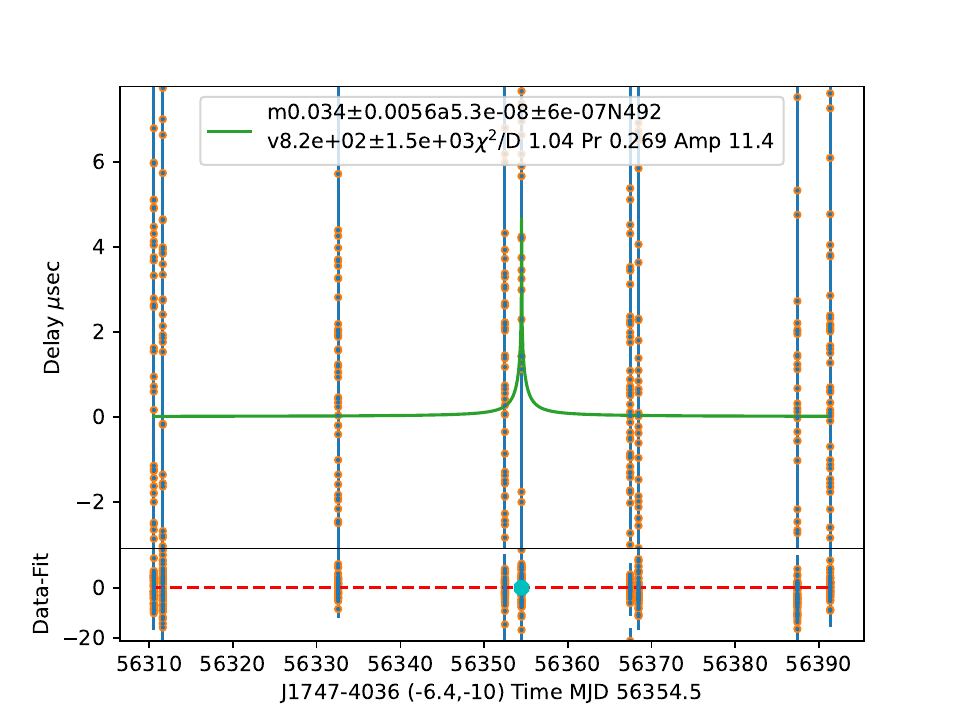}
\includegraphics[width=\columnwidth,height=0.28\paperheight]{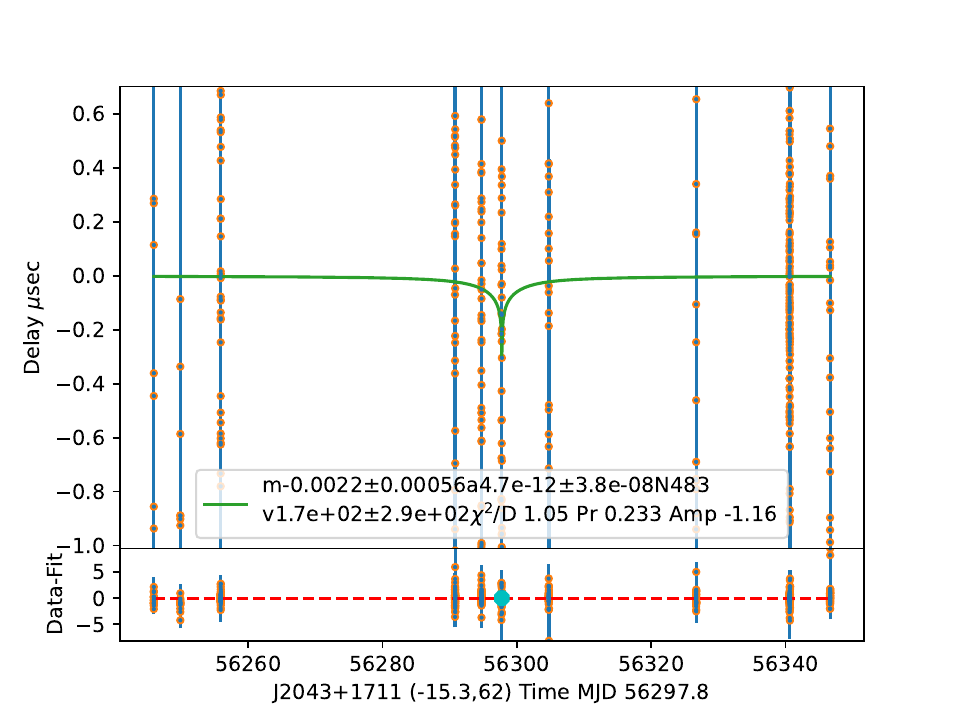}
\caption{\label{fig:Redfits} The fits for the two candidate events found in the red noise subtracted data.}
\end{figure}

\begin{table*}
\caption{\label{tab:results2}There are 2 candidates after duplicates have been removed.
	Two pulsars are represented.  The columns are the same as in Table \ref{tab:results}}
\begin{tabular}{|c|c|c|c|c|c|c|c|c|}
\hline
Pulsar & Sample & Duration & Count & RMS & Significance & $\mathcal{M}\pm\sigma_\mathcal{M}$ in $\mathcal{M}_{Sun}$ & $\mathcal{M}/\sigma_\mathcal{M}$ & MaxAmp \\
\hline
J1747-4036 & 25 & 80.8 & 492 & 6.215 & 0.2691 & 0.03391$\pm$0.00559 & 6.065 & 11.43 \\
J2043+1711 & 13 & 100.7 & 483 & 1.408 & 0.2329 & -0.002218$\pm$0.000565 & -3.925 & -1.155 \\
\hline
\end{tabular}
\end{table*}
\begin{table*}
\caption{\label{tab:summary2}Summary of the fit parameters for the 2 events.  The columns are the same as in Table \ref{tab:summary}}
\begin{tabular}{|c|c|c|c|c|c|}
\hline
Pulsar & Sample & $\mathcal{M}\pm\sigma_{\mathcal{M}}$ & $d\pm\sigma_{d}$ & $v\pm\sigma_{v}$ & $T_{0}\pm\sigma_{T_{0}}$ \\
\hline
J1747-4036 & 25 & 0.03391$\pm$0.00559 & 0.000003011$\pm$0.00003463 & 819.9$\pm$1478 & 56354$\pm$0.00000067269 \\
J2043+1711 & 13 & -0.002218$\pm$0.000565 & 2.688E-10$\pm$0.000002152 & 168.7$\pm$285.1 & 56298$\pm$0.00000037965 \\
\hline
\end{tabular}
\end{table*}
\begin{table*}
\caption{\label{tab:sample2}Summary of the data sample of the 2 pulsars.  The columns are the same as in Table \ref{tab:sample}}
\begin{tabular}{|c|c|c|c|}
\hline
Pulsar & Sample Length (MJD) & Sample Size & Mean Residual \\
\hline
J1747-4036 & 609.33 & 2771 & -8.21E-17 \\
J2043+1711 & 833.75 & 1382 & 5.14e-18 \\
\hline
\end{tabular}
\end{table*}
\section{\label{sec:level2}Timing Noise}
There are known sources of timing noise (also known as red noise) that can mimic a Shapiro delay signature \cite{Goncharov}.  The search for gravitational encounters has been conducted on the
same data sample with additional processing to remove timing (red) noise.  Two candidates remain.  Figure \ref{fig:Redfits} has the plots for these events.
Tables \ref{tab:results2}, \ref{tab:summary2}  and \ref{tab:sample2} have the details.

Pulsar J2043+1711 has a candidate encounter in the red noise subtracted data but the event didn't make
the cut in the dispersion measure sample search.  The event was found in the initial dispersion measure sample
at the same location, with the same mass and velocity and the same $\chi^{2}/DOF$.  It did not make the cut
to proceed to a 10\% symmetric fit because the mass found was only 1.7 times $\sigma_{\mathcal{M}}$.  The searched samples overlap so even when missed in sample 13
it could have been seen in the previous sample.  J2043+1711 sample 12 also
failed to move to the 10\% centered fit by failing the $\mathcal{M}/\sigma_{\mathcal{M}}>2$ cut.

\section{\label{sec:level2}Discussion}
This search has failed to find definitive evidence for a significant occurrence of
gravitational induced delays as modeled in this search in the 65 pulsar IPTA2 data sample.
Four out of nine of the events in Table \ref{tab:results} and one out of two of the events
in the red noise subtracted table \ref{tab:sample2} have $\mathcal{M}<0$ which is unexpected from a lump of matter.  The fact that the number of positive and negative events are similar suggests a common source.

The red noise subtracted candidates have small masses and high velocities which could be due to an unmodeled short period detector time shift.  The referee suggested removing the
epoch at the fitted time and refitting.  This removes the candidate encounters, verifying
the hypothesis.  True signals should span many epochs, if only to reduce susceptibility to
noise.

The relativistic time delay due to masses in the source system or in the solar system
are found and removed.  Why is the common local distribution of matter not manifest in
the data simple, even if most of the gravitational potential, such as that due to dark matter is smooth?

Equation \ref{eqn:ptdmgeom} has a time dependence on $v$.  As seen after equation \ref{eqn:ptdmgeomder} $v$ is a time rate of change of the angular separation between the pulsar and the matter concentration of the encounter, $v=V/Z$.
As such, it decreases with distance.  The search has been conducted for
encounters that return to a baseline of 10\% of the peak.  If $v$ is small the delay
may not return to a baseline value during the duration of the sample.  The absence of a
change in the delay time on the scale of the observations make a gravitational induced delay indistinguishable from other sources of time delays.

Extending the search to lower $v$ is possible.  Most of the pulsars have many years of data.  But increasing the residual sample length introduces the possibility of overlapping
signals which may not fit this simple model.

\section{\label{sec:level2}Conclusions}
This article presents a search for massive objects via
their gravitational fields.  A four parameter fit has failed to identify convincing
sources.  In principle the paramters in the fits were unbounded.  In practice the nature of the data sample imposed bounds.  The RMS set the scale for the amplitude, the duration set a limit on detectable velocities and the sample spacing and short term nose limited
the sensitivity to the impact parameter.

\section{\label{sec:level2}Future Possibilities}
Classic gravitational lensing \cite{Ref64} provides three observables, the time delay used
in this analysis, amplification of the signal used in microlensing and multiple images.  It may be possible to add these to the pulsar observables.
The recorded energy may show changes correlated with the time delays or
one may be able to discern changes in the apparent source size correlated with the delay.  Even modest additional input may help reject noise events
and better characterize true lensing sources.

\section*{Acknowledgments}
This work would not have been possible without the efforts of the maintainers
of the IPTA second data release and the Tempo2 code used to fit the residuals.
Jim Rich has been an invaluable aid in understanding the microlensing method
of observing dark matter that has many concepts in common with those used here.
The author is grateful to an anonymous referee who identified several shortcomings in an earlier version of the manuscript.

\section*{Data Availability}
The data underlying this article is from the IPTA second data
release \cite{IPTA} which is publicly available.  Please see their article
for details on how to access the data sample.  The computer code for this analysis is available from the author.  Tempo2 is available from 1ts authors \cite{Tempo2}.
\bibliographystyle{mnras}

\bsp    
\label{lastpage}

\end{document}